\documentclass[
amsmath, amssymb, 
aps, pra,
reprint, twocolumn,
superscriptaddress,
longbibliography,
floatfix,
showkeys
]{revtex4-1}

\usepackage{graphicx}
\usepackage{csquotes}
\usepackage[urlcolor=blue, colorlinks=true,citecolor=blue]{hyperref}

\usepackage{color}
\usepackage{bbm}
\usepackage{nicefrac}

\definecolor{orange}{RGB}{255,127,0}

\begin{document}
\title{What the foundations of quantum computer science teach us about chemistry}

\author{Jarrod R. McClean}
\email[Corresponding author: ]{jmcclean@google.com}
	\affiliation{Google Quantum AI, 340 Main Street, Venice, CA 90291, USA}
\author{Nicholas C. Rubin}
    \affiliation{Google Quantum AI, 340 Main Street, Venice, CA 90291, USA}
\author{Joonho Lee}
    \affiliation{Google Quantum AI, 340 Main Street, Venice, CA 90291, USA}
    \affiliation{Department of Chemistry, Columbia University, New York, NY 10027, USA}
\author{Matthew P. Harrigan}
	\affiliation{Google Quantum AI, 340 Main Street, Venice, CA 90291, USA}
\author{Thomas E. O'Brien}
	\affiliation{Google Quantum AI, 340 Main Street, Venice, CA 90291, USA}
\author{Ryan Babbush}
	\affiliation{Google Quantum AI, 340 Main Street, Venice, CA 90291, USA}
\author{William J. Huggins}
    \affiliation{Google Quantum AI, 340 Main Street, Venice, CA 90291, USA}	
\author{Hsin-Yuan Huang}
    \affiliation{Institute for Quantum Information and Matter and \\ Department of Computing and Mathematical Sciences, Caltech, Pasadena, CA, USA}
\date{\today}     

\begin{abstract}
With the rapid development of quantum technology, one of the leading applications that has been identified is the simulation of chemistry.  Interestingly, even before full scale quantum computers are available, quantum computer science has exhibited a remarkable string of results that directly impact what is possible in chemical simulation with any computer.  Some of these results even impact our understanding of chemistry in the real world.  In this perspective, we take the position that direct chemical simulation is best understood as a digital experiment.  While on one hand this clarifies the power of quantum computers to extend our reach, it also shows us the limitations of taking such an approach too directly.  Leveraging results that quantum computers cannot outpace the physical world, we build to the controversial stance that some chemical problems are best viewed as problems for which no algorithm can deliver their solution in general, known in computer science as undecidable problems.  This has implications for the predictive power of thermodynamic models and topics like the ergodic hypothesis.  However, we argue that this perspective is not defeatist, but rather helps shed light on the success of existing chemical models like transition state theory, molecular orbital theory, and thermodynamics as models that benefit from data.  We contextualize recent results showing that data-augmented models are more powerful rote simulation. These results help us appreciate the success of traditional chemical theory and anticipate new models learned from experimental data.  Not only can quantum computers provide data for such models, but they can extend the class and power of models that utilize data in fundamental ways.  These discussions culminate in speculation on new ways for quantum computing and chemistry to interact and our perspective on the eventual roles of quantum computers in the future of chemistry.
\end{abstract}

\maketitle
\section{Introduction}
The study of quantum computing in the abstract is an opportunity to ask ourselves what is possible if we could attain an almost unimaginable level of control of the microscopic facets of our universe.  It was first proposed as a solution to the problem of simulating physical systems with strongly quantum characteristics, a task that has proven very challenging for traditional computers~\cite{feynman1982simulating}.  The idea was that if, like the puppet of a marionette puppeteer, one could make a precisely controllable quantum system act enough like a more interesting system, the puppet could answer previously unknown questions about the true system.  This core concept of quantum simulation eventually merged with modern computer science to form the fields of quantum computer science and quantum computing~\cite{deutsch1985quantum}.  This merger allowed these concepts to be made more precise and expanded applications beyond physical systems into abstract ones like breaking cryptography~\cite{shor1994algorithms,ekert1996quantum}.

Despite the expansion into other applications, the simulation of quantum systems, and especially strongly correlated chemistry, has remained a primary application of interest.  Chemistry represents a sweet spot of quantum effects strong enough to make them challenging for classical computers, while still having a well known application space to motivate development.  Since the original proposal of Aspuru-Guzik et al.~\cite{aspuru2005simulated}, there have been a number of developments in quantum algorithms for the direct simulation of chemistry bringing costs down from astronomical numbers to routine calculations for even very challenging systems~\cite{wecker2014gate,mcclean2014exploiting,babbush2016exponentially,babbush2019quantum,google2020hartree}.  

In parallel to algorithmic developments, quantum technology has advanced at a rapid pace.  Recent demonstrations by the Google group have experimentally shown that these devices are capable of tasks that are incredibly challenging on a classical computer, with a gap that will only grow with the quality of modern quantum computers~\cite{arute2019quantum}.  In addition, a number of prototype chemistry experiments have now been experimentally demonstrated on quantum devices~\cite{lanyon2010towards,peruzzo2014variational,o2016scalable,google2020hartree}.  At present, error rates in quantum devices are too great to make them competitive with the best classical algorithms for chemistry, but quantum error correction and mitigation offer paths forward~\cite{calderbank1996good,gottesman1997stabilizer,mcclean2017hybrid,huggins2020virtual,lee2020even}.

\begin{figure}[t!]
\centering
\includegraphics[width=8cm]{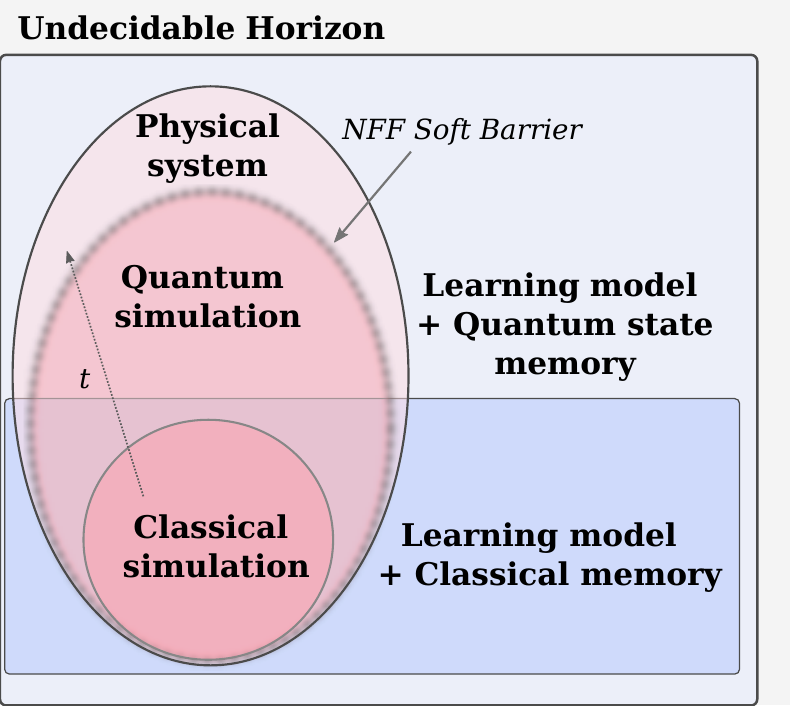}
    \caption{Cartoon of the relative power of direct simulation vs learning models for chemistry, where a point on the figure is a question related to a chemical system.  The ovals on the left depict questions efficiently accessible via simulation of time dynamics, where quantum simulation is widely believed to be exponentially stronger than classical simulation, but still softly bound (non-exponential) by the no-fast-forwarding (NFF) to lag behind the system in nature even with a quantum computer.  Learning models, which include both traditional theories supported by data like thermodynamics as well as modern machine learned models, strictly include all questions answerable by traditional simulation, as they are assumed to have access to simulation but also have their power enhanced by data that comes either from nature or other quantum simulations.  The availability of quantum states held in memory and full quantum computation extends the questions that may be efficiently answerable beyond that of traditional learning models.
    \label{fig:SpeedLimit}
    }
\end{figure}

However, even without an actual quantum computer running an algorithm, the study of quantum computer science in the abstract has led to insights about the limitations and possibilities of chemical simulation, and indeed under modest assumptions, about the universe and chemistry itself.  In this work, we highlight some of these developments leading up to a perspective on the role both traditional and quantum computers may play in the future of chemistry.  To build to this perspective, we begin by framing direct chemical simulation as a digital experiment.  In this framework, we exploit known results that restrict the power of any computer, even a quantum one, to understand limits of such digital experiments in chemistry.  Along the way, we encounter surprising results about the impossibility of an algorithm for determining if a system ever thermalizes, flying in the face of conventional thermodynamic analysis.  However, this construction guides us to new ground, where we will see that learning from data can be fundamentally more powerful than traditional computation.  This sheds new light on the success of many existing chemical theories and the way in which they exceed rote simulation by leveraging the additional power provided by data.  

The relationship between traditional simulation and learning models in chemistry is cartooned in Fig.~\ref{fig:SpeedLimit}, where classical models with data from nature can exceed the physical prediction horizons of rote simulation.  We consider learning models to be models where some amount of high quality training data for the task is available, either from a physical experiment or a simulation in addition to just the specification of the problem.  As we will argue, this is not just a reference to modern machine learning methods, but rather encompasses the foundations of chemical theory such as transition state theory, molecular orbital theory, and thermodynamically controlled reactions.  In addition, we take the opportunity to highlight previously unused technology from quantum information science that may bolster the development of chemistry, even before the arrival of full fault tolerant quantum computers.  The links from chemistry to results in quantum computer science all build to our ultimate perspective that the eventual role of quantum computers in chemistry will be to aid in the construction of learning models for chemistry.  This includes providing data to classical models, constructing quantum models that can make accurate predictions with far less data, and eventually interfacing directly with quantum data from chemical systems.  To close, we wrap this perspective into an outlook for the interplay between these two exciting areas.

\section{Digital chemistry experiments with quantum computers}
The idea to draw a distinction between theory and computation is not new, with suggestions of considering computation as the ``third pillar of science'', alongside experiment and theory~\cite{reed2005computational}.  In this framework, it can sometimes be more accurate to view simulations as closer to experiments than to theory.  Quantum computing, and especially quantum simulation of physical systems, helps to make this even more clear by constructing simulations of the physical world at the quantum level with efficiently refineable (and bounded) accuracy.  In addition, limitations on precision for quantum computers~\cite{Knill:2007} to learn physical properties sometimes place them even closer to experiments than classical digital simulations of the same systems.

Indeed, the most natural setting for a quantum simulation on a quantum computer is that of watching the system step forward in time, or quantum time dynamics.  Within the field of quantum computing, the term ``quantum simulation'' is sometimes reserved for time dynamics simulation specifically, and when we refer to rote simulation, it will typically apply to this case.  That is, one sets an initial state, a physical system defined by its interactions, and some time to simulate, and the quantum computer performs a computational experiment in mapping the initial state to the final time, where any reasonable observable of the system can be probed. This time dynamics obviously mirrors the most natural settings of the real world.  However, despite this natural setting, one will notice that a large number of quantum algorithms and theories closely follow that of electronic structure performed on classical computers, where the focus is on low energy eigenstates of the electronic system~\cite{aspuru2005simulated,helgaker2014molecular}, rather than exclusively the final state after some fixed evolution.  This important deviation already wraps in data from the natural world that tells us in many systems, thermal states are an apt description of a system and at low temperatures low energy states are most commonly observed.  As we will discuss later, the use of such observational knowledge is strictly more powerful than time dynamics simulation, and underpins the great success of many modern chemical theory or simulation methods.  A detailed discussion of this point will follow, but for now we lump both static and dynamic experiments into the category of direct simulation.

By direct simulation, we mean a simulation where the system of interest (often processes such as a reaction) is known, and we seek to use the computational experiment to gain insight into the process that is challenging to access through experiment.  For example details of a reaction mechanism, transition states, or migration of charge through a system.  They represent the bulk of computational experiments, both on classical computers, and those proposed on quantum computers, and we draw a distinction between direct simulation and design, often characterized as inverse problems~\cite{sanchez2018inverse}.

For the task of direct simulation of chemical systems, quantum computers have been shown to demonstrate an exponential advantage over their classical counterparts under modest assumptions~\cite{aspuru2005simulated,kassal2008polynomial,cao2019quantum,mcardle2020quantum}.  An exponential speedup has the practical implication that some simulations that might have taken longer than the age of the universe, could be done in mere seconds.  More specifically, full quantum dynamics with no approximations beyond discretization error can be performed on a quantum computer in a time that scales only polynomially in $t$, $N$, and $M$ where $t$ is the simulated time, $M$ is the number of basis functions and $N$ is the number of electrons~\cite{babbush2019quantum,von2020quantum,lee2020even}. Interestingly, along with these speedups it comes with some of the same limitations possessed by a physical experiment. For example, in contrast to classical simulations where more precision in a simulated quantity is often relatively easily obtained, the Heisenberg limit applies to any measurements one might perform to extract information. These approaches have also included exploring ansatz for electronic systems that are inaccessible to classical computers~\cite{peruzzo2014variational,mcclean2016theory}.  Perhaps surprisingly, recent results have shown that quantum computers can even achieve a scaling for exact computations that is sub-linear in the basis set size, by taking advantage of a first quantized representation in a way that has no current classical counterpart~\cite{babbush2019quantum}.

However, quantum computers do not vastly extend their capabilities into the realm of discovery or design, which we identify as distinct from direct simulation.  That is, while many of the proposed simulations would be faster and more accurate representations of the same systems than would be achievable classically, they are much the same type of experiment without reaching into the design space.  While there has been some notion of how quantum search may assist in design, most results promise at most a quadratic speedup in contrast to exponential speedups in direct simulation~\cite{barkoutsos2021quantum}, though we note the direct simulation subroutines used in a potential search would still benefit from improved speed or accuracy.  Due to the immense size of the design space, this quadratic speedup is not terribly compelling if it is not combined with structured strategies.  That is, any improved search must start not from naive global search, but a search enriched by knowledge of the design space.  For example one that starts in the chemical neighborhood of known, synthesizable compounds.  In addition, practical issues arising in real quantum computers may prevent quadratic speedups from being advantageous for some time~\cite{babbush2020focus}.  We will argue later that perhaps it is even more apt and useful to consider an infinite chemical space as opposed to one which is merely combinatorially large.

That said, direct simulation, of course, has great value for design.  For example, if a key reaction or catalyst is known to work to catalyze a reaction of interest, the microscopic detail of a computational experiment that is inaccessible in the lab can offer insight into the causal mechanism that enable the design of analogs~\cite{norskov2009towards,ess2019introduction, knizia2013intrinsic}.  In addition, a now popular approach is the use of screening of many candidate compounds through direct simulation in order to identify new candidate molecules or learn properties which are predictive for a desired design task~\cite{hachmann2011harvard,jain2013commentary,singh2015computational}.  This type of approach is also used for selection of candidates in protein to small-molecule docking~\cite{kinnings2011machine}. Direct simulation can also be used for novel discovery of processes through direct time evolution, as has been done with the molecular Nano-Reactor~\cite{wang2014discovering}.  Quantum computers serve to strictly improve this type of discovery by expanding the set of systems that can be accurately treated in a given amount of time by such methods.   Advantage in improving the accuracy in direct simulation has been the primary focus in much of the literature of quantum computation for chemistry, with the hope that it enables accurate study of important systems inaccessible to current methods.  The example receiving perhaps the most attention is FeMoCo~\cite{reiher2017elucidating,motta2018low,cao2019quantum,berry2019qubitization,von2020quantum,lee2020even}, where the hope is that direct simulation will offer mechanistic insight that leads to the design of new catalysts for nitrogen fixation.  However, when leveraging direct simulations results for improved design, the simulations are typically tightly coupled to a theory already rooted in data.  For example, the energy of a transition state might be used to justify a reaction rate, but this is inherently coupled to the empirical success of transition state theory, rather than a result which emerges directly from simulation.

In contrast, comparatively little attention has been paid to the potential advantages of improved dynamics simulations, perhaps due to the larger foundation of work available utilizing static eigenstates in the classical case.  In addition, it is perhaps surprising that dynamics simulations on quantum computers that fully relax the Born-Oppenheimer approximation, using fully quantum nuclei and exact electronic representations, are about as costly as those that live within such approximations.  However, with each step towards this nuclear-electron soup, what is gained in accuracy might be paid in ease of interpretation as the added overhead of re-discovering the notion of a molecule through observable measurements remains a largely uncharted problem in the context of quantum computing~\cite{bochevarov2004electron,kassal2008polynomial}.  Taking advantage of these departures from the tried and true anchors of chemical space, and embracing the idea of a fully \textit{ab initio} experiment will require more development in how this type of simulation can be most beneficial.  However, if it is unlocked, it may offer new insights into complex dynamic processes in spectroscopic or electrochemical processes where both excited electronic and quantum nuclear states can play a role~\cite{wang2014discovering,martinez2017ab}.  It may also offer new insights into the long studied challenges surrounding proton coupled electron transfer~\cite{weinberg2012proton}.

Given the immense power of quantum computers in speeding up these types of direct simulation, one is often left wondering how far these advantages extend.  This question is one that has been studied in great detail in computer science, and we now wish to turn to our understanding of the limits of power in quantum computation.  The similarity between quantum computations and the physical world imply that the results also tell us something about nature and chemistry itself, and we now turn to this perspective. 

\section{Limits of even quantum computations in chemistry}
Having set up the perspective of quantum computation as digital experiments, we now turn to the implications this may have for the physical systems themselves.  Due, in part, to the challenge of building scalable quantum hardware, for a good part of its history quantum computing was a largely theoretical endeavor that aimed to understand the power of quantum devices, and by extension, physical systems.  Considerable progress has been made on this problem, and we wish to highlight those results we believe have the most bearing on chemistry.

Before discussing how they pertain to chemistry, it is worth mentioning a few aspects of the assumptions made in stating the results below.  In particular, in order to talk about scaling and cost, it is necessary to have some model of computation where cost can be quantified.  We will generally be assuming this means we use a finite, often local, set of operations called quantum gates, and we work within a digital gate model of quantum computation.  This assumption is in some ways analogous to assuming nature favors local interactions or that computers are built from modular components.  Moreover, some of the results below most precisely pertain to the number of times some information, such as the Hamiltonian of a system, must be accessed.  This is known as the query complexity, and while it is often closely related to time complexity, these can be different in some circumstances, such as when parallelization is or isn't possible and we attempt to be clear as to when this distinction is important.

\subsection{Quantum computers can't fast-forward time either}
Starting with the most direct form of quantum simulation, time dynamics of a quantum system, we have already noted the exponential advantage over classical simulation achieved by current quantum algorithms.  This begs the question how far these quantum algorithms can go in accelerating simulation of physical systems.  For example, can a quantum computer simulate a system faster than nature itself evolves the same system, that is, sub-linear in the time being simulated?  It turns out that, assuming one only has the available space to represent the system serially, the answer to this is a resounding no.  It is also worth noting that if a system is finite and the time to simulate is larger than the number of possible system states, the cost of diagonalization, finite state enumeration, or some equivalent can be paid to fast-forward a system.  However, in most cases the time-scale of interest is much less than this, and indeed as we argue later, infinite perspectives on chemical systems are sometimes more natural.

While there are some special systems that can be fast-forwarded~\cite{gu2021fastforwarding}, it cannot be done in the most general case.  This result can be understood at a high level by imagining if this were true and realizing one could build a recursive simulation that does any computation in vanishingly small time, by nesting simulations within simulations with no space overhead.  If one allows the space on the quantum computer to expand polynomially with the simulation time, this result is more unclear and related to the classical debate of if any computation can be parallelized arbitrarily well, encapsulated by the theoretical question of if P=NC~\cite{greenlaw1995limits}.  This is an open question that currently holds a similar status to the more famous question of if P=NP.  Some progress has been made on the question if all computations can be parallelized arbitrarily for specifically quantum computers, showing that more simulation time is strictly more powerful relative to an oracle~\cite{coudron2020computations}.  To simplify the message here, we assume that quantum space resources are the minimum required to perform the simulation, and that as a result oracle queries must be done in serial. More formally, the limitations of simulation are encapsulated in the no-fast-forwarding theorem~\cite{berry2007efficient,atia2017fast}.  This theorem importantly shows that not only is universal computation not fast-forwardable, but surprisingly even some Hamiltonians that are incapable of universal computation cannot be fast-forwarded either.

Under the stated assumptions, the no-fast-forwarding result practically implies that no simulation which retains a complete level of detail at the quantum level can simulate physical systems faster than nature, and likely there will be some large factor disadvantage between them.  This result is in some ways not unexpected from classical simulations of classical chemical models like molecular dynamics, where for a fully detailed tracking of the model in the most general case, one expects to lag behind the physical timescale of the system unless approximations or reduced models are utilized. The emphasis here is that this limitation provably carries into the case of using a quantum computer as well. Taking this result at face value along with the knowledge that important chemical reactions like rusting can take months to occur, using a single molecular realization in a simulation could take comparable times to witness a reaction event, making discovery in this fashion totally impractical.  However, most chemists would know that is not the end of the story. 

Specifically, methods like transition state theory based on static calculations of free energies or transition path sampling ~\cite{dellago1998transition} can allow access to rate constants on time scales much faster than direct physical simulations~\cite{phillips2005scalable}.  Hence, while the above theorem is true for precise simulations of the full quantum state, we see from practice that physical theories grounded in observation can circumvent such wild overheads by eliminating the complexity of the full quantum state and focusing on a reduced set of observables, reduced precision, or approximate models amenable to fast-forwarding.  Such reductions of the number of important degrees of freedom have been studied in various contexts to try to understand ``what makes science possible''~\cite{machta2013parameter,transtrum2015perspective}. It should also be noted that these results are about the most general system, and specific systems may exhibit physics that allow faster than physical simulation when that data is utilized by validating approximate models.  In other cases, particular models are amenable to classical fast-forwarding, like non-interacting electron models, but they are not general enough to encompass all quantum systems with interactions.

The observation that chemical methods have been so successful in making predictions beyond physical time scales leaves one believing that perhaps physical systems always admit some form of reduction that will allow us to forecast results that we actually care about (as opposed to a full quantum state), well in advance of simply stepping them forward in time.  A recent string of results in quantum computer science has proven that, in fact, even in some very simple systems by construction, there must exist some questions that are fundamentally unanswerable ahead of time.  This list of problems surprisingly includes whether simple one-dimensional (1D) systems thermalize~\cite{shiraishi2020undecidability}, which has implications for the predictive power of thermodynamics in general systems.  These results are much stronger than simple chaos of dynamic systems and require some exposition, but we believe they are a value lens through which to view chemical problems.

\subsection{Questions no algorithm can answer - undecidability}
A direct reading of the no-fast-forwarding theorem contrasted with the unassailable predictive success of simplified chemical models raises a number of interesting questions.  On the one hand, the no-fast-forwarding theorem tells us that even with an exponential advantage over classical simulation of quantum dynamics, a quantum computer aiming to simulate the dynamics of a full wavefunction to a fixed accuracy is bound by some multiplier of the physical time of the actual process.  On the other hand, we see that reduced models, for example mean-field molecular orbital models, reductions based on evaluating the energy of stationary states like ground and excited states, or even arrow pushing in Lewis dot structures can sometimes be powerfully predictive for chemical phenomenon of interest at a relative time cost much less than the time scale of a process of interest.  This leads one to ask if there is always a predictive model or algorithm for simplified questions.   Surprisingly, results in computer science have shown us that there are indeed even relatively simple questions related to physical systems for which it can be proven that no single algorithm can answer for all such systems in finite time.  Here, while we will certainly not claim that this should cause one to abandon current methods, we will argue that these results should have implications for how one approaches some chemical problems.

To make our case, we will start with a motivating example before we move to the more general concepts that will allow us to address some of the tensions with this interpretation.  Consider the question of ``Given unbounded time, does this physical system reach thermal equilibrium?''.  The basic assumptions of conventional thermodynamic theory, leaning on the ergodic hypothesis~\cite{neumann1932proof,kennard1938kinetic,boltzmann2012lectures}, assert that the answer is a resounding ``yes'' under assumptions that the system is ergodic.  Indeed many standard molecular dynamics simulations exploit both possible directions, either determining free energies through proxies derived from long time simulations~\cite{phillips2005scalable} or predicting long time behavior through models for free energies. However, recent results have shown that there are 1D translationally invariant systems for which no algorithm that can answer the question of thermalization or even simple questions about energy gaps in finite time~\cite{cubitt2015undecidability,humphreys2015more,bausch2020undecidability,shiraishi2020undecidability}.  

To make such a problem even more concrete, let's take the question of a finite gap as an example. The problem statement can be setup briefly as follows.  One is provided as input a finite Hamiltonian for a chunk of a system that will be repeated indefinitely.  While any finite chunk of this system could be diagonalized, determining the existence of a gap for that chunk, this proof implies that no finite number of chunks taken together could ever tell you about the status of the gap for the infinite system in the general case.  Moreover, any extrapolation on those chunks to a large size limit cannot be guaranteed to converge.  The statement is so strong that we can guarantee that not only will no simple algorithm work, but there cannot exist such an algorithm of any finite time complexity.  This type of problem is known as an undecidable problem, the first example of which was the famous halting problem of Turing~\cite{turing1936halting}.  This class of problems is much more challenging than even the perhaps more familiar class of NP-complete problems, as even when provided with a supposed answer, a mortal prover cannot verify if it is correct.  This implies that for a number like an energy or free-energy representing a system, either that number cannot be generally predictive of the behavior of that system, or the number itself must be uncomputable.  Showing that this is true, even for a quantum computer, which can perform classical computation as a special case, underscores the weight of these results.

The idea that bare thermodynamic quantities may not be predictive due to lack of thermalization is certainly not new in the study of non-ergodic systems, with preeminent examples like spontaneous symmetry breaking, spin-glasses, and the general idea of a Gibbs measure~\cite{palmer1982broken,chikazumi2009physics,prigogine1967symmetry,georgii2011gibbs}.  These results strengthen these concepts to tell us that in some non-ergodic systems, there is no computable replacement for something like free-energy that is predictive of the behavior of a system.  Hastily defined, the halting problem is the problem of finding an algorithm that, given a finite program input, can determine in finite time if the program halts or runs forever.  If one views a physical evolution as a computation where halting is defined as some qualitative change~\cite{moore1990unpredictability}, for example, the fraction of random systems which thermalize should be closely related to the fraction of random programs that halt, a number known as Chaitin's constant~\cite{chaitin1975theory}.  Indeed, it is easy to see such a thing requires solving the halting problem for every possible instance to determine the digits of the fraction of halting programs, which defines it as an uncomputable number.  For such systems, assuming ergodicity is analogous to solving the halting problem by simply hypothesizing that all programs encoded by the physical evolution halt.

\begin{figure}[t!]
\centering
\includegraphics[width=0.83\columnwidth]{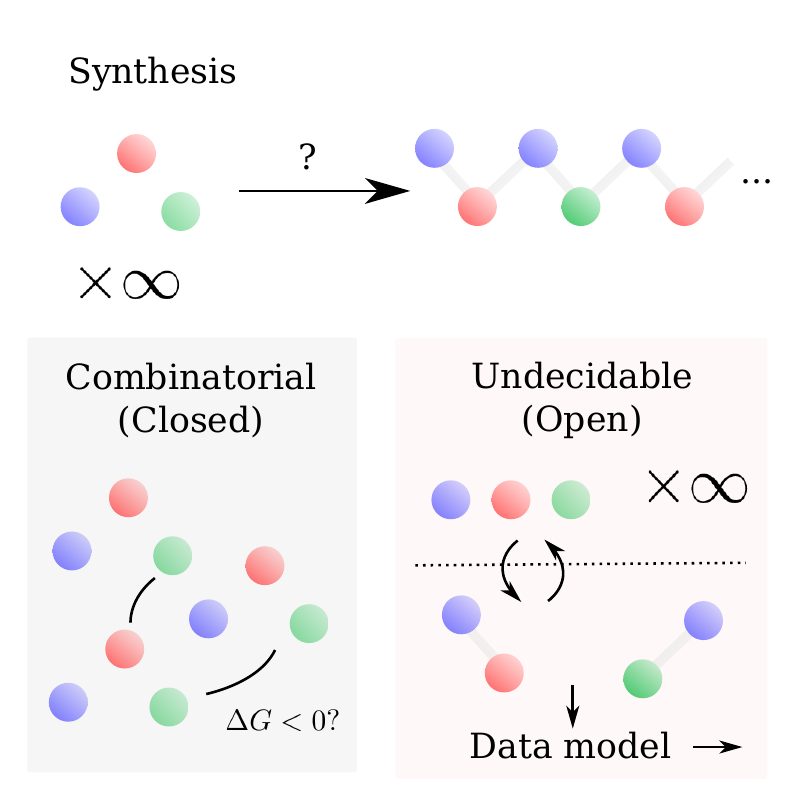}
    \caption{ Sketch of two viewpoints for the problem of chemical synthesis.  In this scenario, one wants to know if a certain target compound can be reached with a feedstock of specified chemicals under typical conditions.  The combinatorial view tries to address this question by introducing a closed reaction system with a large but finite number of the reagents, and explores potential reaction pathways to determine feasibility through some proxy like free energy $(\Delta G)$ or a short time simulation.  The undecidable view point embraces an open systems picture and the relationship to DNA computing to argue that the most feasible approach tries to use data either from nature or related simulations to determine the answer to the question.  Such an approach can more naturally include the effects of common side-reactions or strong kinetic control.
    \label{fig:CombinatorialVsUndecidable}
    }
\end{figure}

For veteran chemists, another way of viewing this result is that it is a vast strengthening of the concept of kinetically controlled reaction systems to be contrasted with thermodynamically controlled systems.   Previous work in this area has shown that this type of unpredictability is distinct and stronger than traditional notions sensitivity induced chaos in rendering systems unpredictable, even when they are deterministic.  Phrased in a physical point of view, for a system exhibiting undecidable dynamics, there are sudden, qualitative changes can occur that cannot be predicted in any way other than simply stepping the system forward in time~\cite{moore1990unpredictability}.  For chemical practitioners, such a description is certainly reminiscent of the emergent phenomena that transition us from simple molecules to proteins, RNA, DNA, complex catalytic networks, and even living systems. 

Before continuing, we must address a tension that arises with the mapping of physical processes to halting problems in the form of finite versus infinite systems.  In order to map to a halting problem rigorously, one typically maps the operations of a physical system to operations of a Turing machine, with an infinite sized tape.  If the tape were finite, one can spend a very long, but not infinite, time examining all possible states of the system to arrive at an answer to any question related to the state of the system.  For example, in studying the dynamics of a finite physical system, one could discretize the possible states, and expend an exponential amount of time in the system size determining an answer.  Hence one might want to argue that since the universe is often presumed finite, these conclusions are of little consequence, even though from a practical point of view, the exponential of even a modest sized finite system would take longer than the age of the universe.  Despite their somewhat similar predictions in practice (too long to compute), we argue that embracing the viewpoint of undecidable dynamics allows one to let go of the impossible dream of treating chemical questions by the exponential enumeration of the combinatorial perspective. Whether or not a physical system exhibits truly undecidable behaviour, embracing the possibility leads to a more natural description of many chemical processes and points the way towards more apt solutions.

We depict an example of this perspective in Fig.~\ref{fig:CombinatorialVsUndecidable}.   Consider a collection of small molecules attached to a reservoir of source chemicals, and we wish to know if a particular strand of DNA will ever form given this setup.  If one took the closed system, or combinatorial, viewpoint, it would suggest that perhaps the most effective way to address this question is to model the reservoir with some enormous, but finite, set of explicit molecules, and check all of the possible states of the dynamic system.  Along the way, one can check things like the change in free-energy ($\Delta G$) or simply proceed with short time dynamics.  Not only is the approach untenable due to the number of possible configurations, but given the nature of self-catalytic reaction networks, one could accidentally include slightly too few molecules to form the right catalytic process to enable the formation of that strand and reach the wrong conclusion.  In contrast, the open system, or undecidable, interpretation of this setup, is that the physical system can continue to borrow from the infinite reservoir like states of the Turing tape.  If it has to temporarily expand to chemical networks of unforeseen size, it can do so.  In this situation, one tends to lean on data models for what has existed before, and we detail later how doing so changes the power of the simulation approach.  The undecidable interpretation is supported by recent work showing that DNA computing is Turing complete and as such can encode the halting problem~\cite{freund1999dna,paun2005dna}.  While DNA computing is a bit of an artificial construction, and real DNA that tends to form in live systems is subject to other constraints that make its activity more predictable, the possibility of implementing universal computation in the dynamics of a chemical system pushes us to embrace the wildest consequences of the theory of computation.

This tension between finite and infinite models is not unfamiliar when related to the study of other non-ergodic systems, like spin-glasses~\cite{palmer1982broken}.  In such cases, models might predict that finite systems have some chance of exhibiting substantial rearrangements, albeit on exponentially long timescales.  However, it has been seen to be useful to lean on a model that takes a limit where these different basins qualitative fracture from each other, and are considered separate.  We suggest here that as molecular systems grow in complexity, the run-away reaction systems that form are best treated by considering them as fracturing into novel parts of chemical space, without tracking vanishingly unlikely events that transition one between these parts of space.  Indeed, the undecidability of thermalization suggests that for infinite systems, it is correct to view such fracturing as exact in the most general case.  Even in simpler cases, this model tracks more closely with the experience of laboratory chemists who have to contend with side reactions rendering a thermodynamically favorable synthesis seemingly impossible. By not splitting hairs about the distinction between a large finite system and a truly infinite one, we can move towards practical alternatives to the brute-force combinatorial approach of pure computation.

But if one resigns themselves to accept that many of these problems have no general algorithm, is one left with no option but to plod forward with rote dynamic simulation or give up?  At least the combinatorial approach offered a glimmer of hope, even if it required longer than the age of the universe for tiny problems right?  In the next section, we argue that in fact this realization does offer an alternative, which focuses the scientist on an approach based on learning and reduced chemical model construction.

\section{Learning as an alternative to computation}
Rather than claim one must give up hope on problems where the most natural formulation looks undecidable or hopelessly expensive, we argue that embracing the undecidable interpretations of these problems frees one from the distraction of the combinatorial approaches, and leads them to embracing the only known resolutions to problems that exist for halting problems.  The first, and not really palatable solution, that has been mentioned, is running time dynamics forward and hoping for the event of interest to occur.  The second, and more actionable viewpoint, is to understand that systems with advice can formally resolve such problems.  While advice has a precise definition in theoretical computer science~\cite{arora2009computational}, it was recently shown that data from the real world can act as a restricted form of advice~\cite{huang2020power}.  To solve halting problems, one's advice would need to constitute a form of infinite time pre-computation, however much like the argument of finite versus infinite systems, the pre-computation performed by the physical universe before this point, though finite, is already quite strong and available to be revealed through experiments.  

To see an example of this in chemistry, one can look to the field of natural product synthesis, where one seeks to synthesize chemical species found in nature from simpler, readily available components in a finite number of steps.  For the most general complex molecule, it can be quite difficult to say if a product can be reached with a fixed set of reagents and conditions or at all, perhaps due to some intervening side reaction.  In fact, we conjecture this form of the problem may be undecidable, since we could specify the target molecule as a particular strand of DNA~\cite{paun2005dna}.  By contrast, in the study of natural product synthesis, the challenging question of chemical synthesizability is answered by nature itself, and practitioners are freed to discover practical routes towards a product rather than ponder if success was even possible theoretically. In this sense, empowering ourselves with observations from the physical world can enable answers to questions that cannot be practically resolved with rote algorithms.  More generally, we believe one could view the models and study of synthesis in organic chemistry and their predictive success in this light.  Hence, taking this point of view suggests that approaches based on learning are fundamentally different than those that rely on computation alone.

With the rise of popularity in machine learning, naturally there is now a flurry of work applying techniques of machine learning to chemistry~\cite{von2020retrospective}.  The applications range from prediction of direct simulation quantities~\cite{gilmer2017neural} to synthesizability~\cite{aykol2019network} and inverse design~\cite{sanchez2018inverse}.  We believe the results we highlight here bolster the motivation for pursuing this line of work in the general sense, but also supports the long standing tradition of chemists developing physical models as well.  While much practical work remains to be done in finding the best representation or approaches, it is now understood that these approaches are fundamentally different, and indeed more powerful in some ways, than traditional simulation approaches.

\begin{figure}[t!]
\centering
\includegraphics[width=0.83\columnwidth]{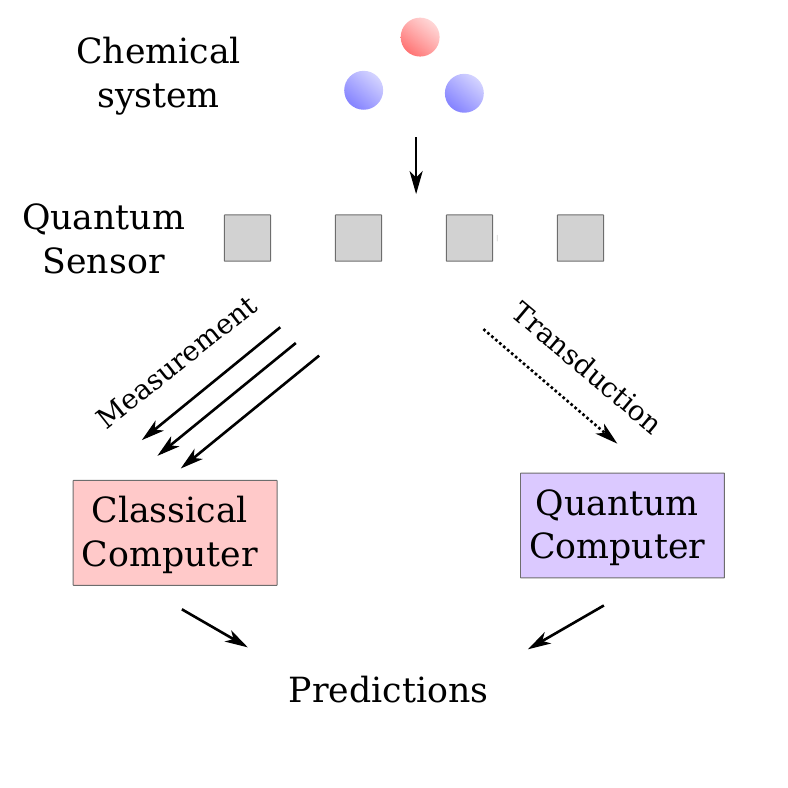}
    \caption{ Comparison of quantum and classical data pipelines.  The presence of classical data or traditional measurements, e.g. data from laser spectroscopy or thermodynamic models, already confers an advantage over traditional simulation.  Here we contrast that pipeline with one were quantum data is collected from quantum sensors, which can be stored in quantum memory, increasing yet again the power of available data.  In the classical case, a sensor measures and reports classical values back to a classical computer.  In the quantum case, a sensor has its state transferred into a quantum computer, a process also known as transduction, where it may be computed alongside multiple additional copies of such states to perform tasks that would be exponentially more costly on a classical computer.  In such setups, the quantum transduction route can require exponentially fewer measurements than the classical case to learn some quantities or perform some tasks.  In cases where data is hard to come by, e.g. transient, short-lived, or rare systems, this can make an impossible characterization into one that is routine.
    \label{fig:QuantumData}
    }
\end{figure}

In particular, recent work has shown that the power offered by data from a quantum computer can lift classical models to be competitive with their quantum counterparts in some cases~\cite{huang2020power}.  If nature is indeed a universal quantum computer~\cite{aaronson2005guest}, this would suggest that quantum computations could help fill the role of natural experiments in providing data to empower learning models, but with improved programmability and flexibility.  This might suggest that in the future, a key role of quantum computers, which may still remain scarce compared to their traditional counterparts, may be to provide data for learning models primarily run on traditional computers.  However, that would be a rather unexciting fate for a technology that pushes the limits of our understanding of the universe.

Instead of accepting this fate, we appeal to other recent work that has shown an overwhelming advantage for quantum computers in how they can process data moved directly into the computer from a quantum sensor, a process referred to as transduction~\cite{huang2021information, aharonov2021quantum}.  A comparison with a traditional data pipeline is shown in Fig.~\ref{fig:QuantumData}.  In such cases, even performing quantum data processing on even very few copies stored in quantum memory can allow one to extract properties of quantum systems that would require exponentially more data in the classical case~\cite{huang2021information}.  Even early machines may be able to manipulate few copies for limited times, and store this quantum data for future models.  This separation is stronger than a traditional computational separation, as if limited measurements are available due to the transient nature of a system, no amount of computation can allow traditional measurement and computation to catch up.  This implies that even systems with a number of qubits that can be classically simulated could prove advantageous.  If the quantum computer has quantum memory that can maintain states in quantum form, this extends the power of these approaches even further.  The relationship between this hierarchy of data models and traditional simulation is depicted in Fig.~\ref{fig:SpeedLimit}. Moreover, if one ventures further into the realm of speculation, recent computer science results have shown that if one can entangle two parties trying to prove something to another, what they can convincingly prove expands to unimaginable heights, including halting problems.  While this celebrated result, MIP*=RE~\cite{ji2020mip}, has not yet been directly connected with the physical world, it is tantalizing to imagine how it might reflect on the power of physical experiments empowered by entanglement to reveal the mysteries of the universe. The question, of course, becomes where to find such useful quantum data and how to effectively get it into the quantum computer.

Chemistry has long benefited from the power of quantum sensors in nuclear magnetic resonance experiments.  While the most basic forms of these experiments are based on ensemble measurements at an effectively high temperature, they offer incredible insight and are fundamentally quantum.  Interestingly, the form of these basic experiments can be closely identified with the so-called one clean qubit model, or DQC1 and this has inspired some recent work in pursuing advantages with NMR~\cite{sels2020quantum}.  However, as NMR technology advances with techniques like hyper-polarization~\cite{leawoods2001hyperpolarized}, multi-dimensional methods~\cite{ernst1966application}, spatial resolution~\cite{devience2015nanoscale}, and zero-field techniques~\cite{zax1985zero}, we appear to be accessing more and more pure quantum states with increasing control~\cite{xia2020demonstration}. In addition the design of more advanced molecular quantum sensors is an active research area~\cite{chung:2021}. With sufficient advances in transduction techniques and developments in quantum error correction, it may become possible to load data from molecules directly into quantum computers, and perform manipulations that are provably challenging for a classical device even with unbounded compute time to replicate, due to the query separations we mention above.  This could enable faster chemical identification, unprecedented accuracy in examining quantum effects in molecules, or fundamentally new techniques for the control and manipulation of chemical states.

\section{Untapped resources in quantum computer science}

The discussion thus far has largely centered around the role of quantum computer science or quantum computers themselves in new approaches to understanding chemistry.  However, it is also interesting to ask more generally, what theoretical tools from quantum information science with potential to impact computational chemistry remain untapped.  For example, techniques in tensor networks that received much of their theoretical development in quantum information, have now begun to inform ansatz for correlated wavefunctions.  Both matrix product states and tensor network states coupled with an entanglement perspective on strong correlation has lead to a wealth of theoretical developments~\cite{sharma2014low, barcza2011quantum}.

\begin{figure}[t!]
\centering
\includegraphics[width=0.83\columnwidth]{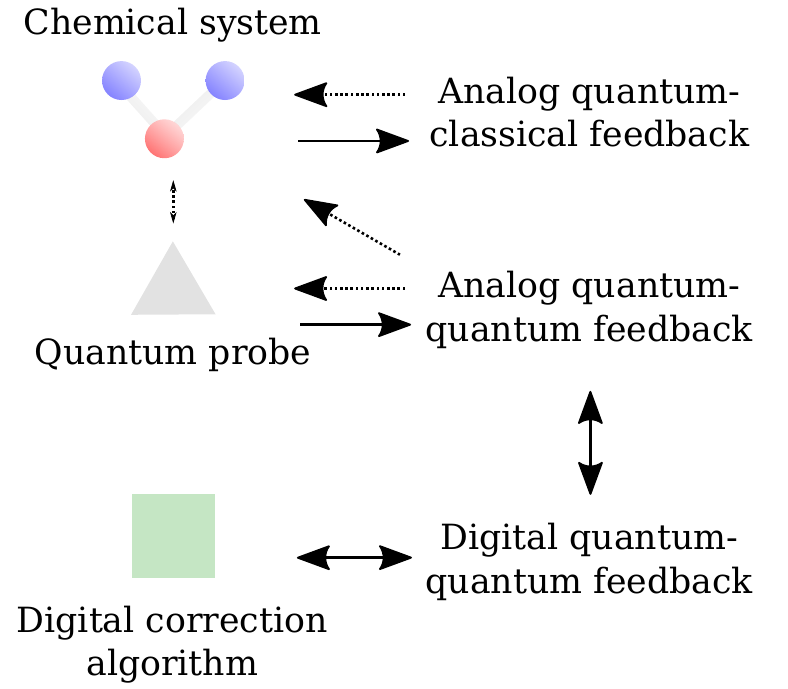}
    \caption{ Digital quantum feedback control inspired by quantum error correction.  Quantum control of chemical reactions is a well studied field, and many aspects of it are mirrored in the control of qubit systems.  While direct measurement of a system through typical analog signals destroys entanglement in a quantum system (quantum-classical), the use of a quantum probe or ancilla system can allow measurements that preserve or create entanglement in the target system reliably (quantum-quantum).  While this paradigm is already powerful, the tools of discrete computer science allow one to use such interactions in combinations with algorithms that can be imagined as sophisticated Maxwell's demons to pump entropy out of the system at incredible rates far exceeding those of typical thermalization to stabilize exotic quantum states, like electronic states, for time scales that could in principle extend for years, using only local measurements and feedback.  These tools are developed at length in the field of quantum error correction, and we imagine here how those tools might eventually integrate with chemical systems.  Such constructions may pave the wave for novel excitonic or energy transport designs.
    \label{fig:Digital_Control}
    }
\end{figure}

An area of remarkable theoretical development that remains relatively unknown in the chemical community is that of quantum error correction~\cite{gottesman1997stabilizer, PhysRevLett.106.110501}.  Perhaps for a chemist, the best way to understand these methods is in relation to the existing and well developed body of research of quantum control of chemical reactions and chemistry~\cite{rabitz2000whither,brif2010control,shapiro1997quantum}.  We depict aspects of this relationship in Fig. ~\ref{fig:Digital_Control}. In quantum control, a signal, often through a laser excitation or similar, may be used to control and measure the system through feedback, however these measurements are often destructive with regards to quantum states in the system, and we refer to this as analog quantum-classical feedback control.  Through the use of a separate quantum probe or ancilla system, one can make measurements of a system that actually generate or stabilize entanglement in the quantum system, or an analog quantum-quantum scheme.  Taking this one step further, one may use multiple probes and digital algorithms that adaptively change based on measurement results to pump entropy out much faster than traditional thermalization would allow.  It turns out the exponentially long suppression of transport out of collection of desired quantum states is reliant on an abstract digital perspective of the states.  These constructions are formalized in the theory of digital quantum error correction, where it has been shown that with only spatially local measurements and classical computation, it can be possible to stabilize exotic quantum states for essentially arbitrary lengths of time with asymptotically modest resources.  If such techniques could be ported more directly into chemistry, stabilizing electronic or excitonic states could lead to novel reactive methods or wildly improved energy transfer efficiencies.  While this remains in the realm of pure speculation, perhaps probe systems that can be interacted with via future versions of tunneling microscopes with near single electron resolution coupled to catalytically active sites on metal surfaces could offer some of the first places to explore these ideas.

\begin{figure}[t!]
\centering
\includegraphics[width=0.83\columnwidth]{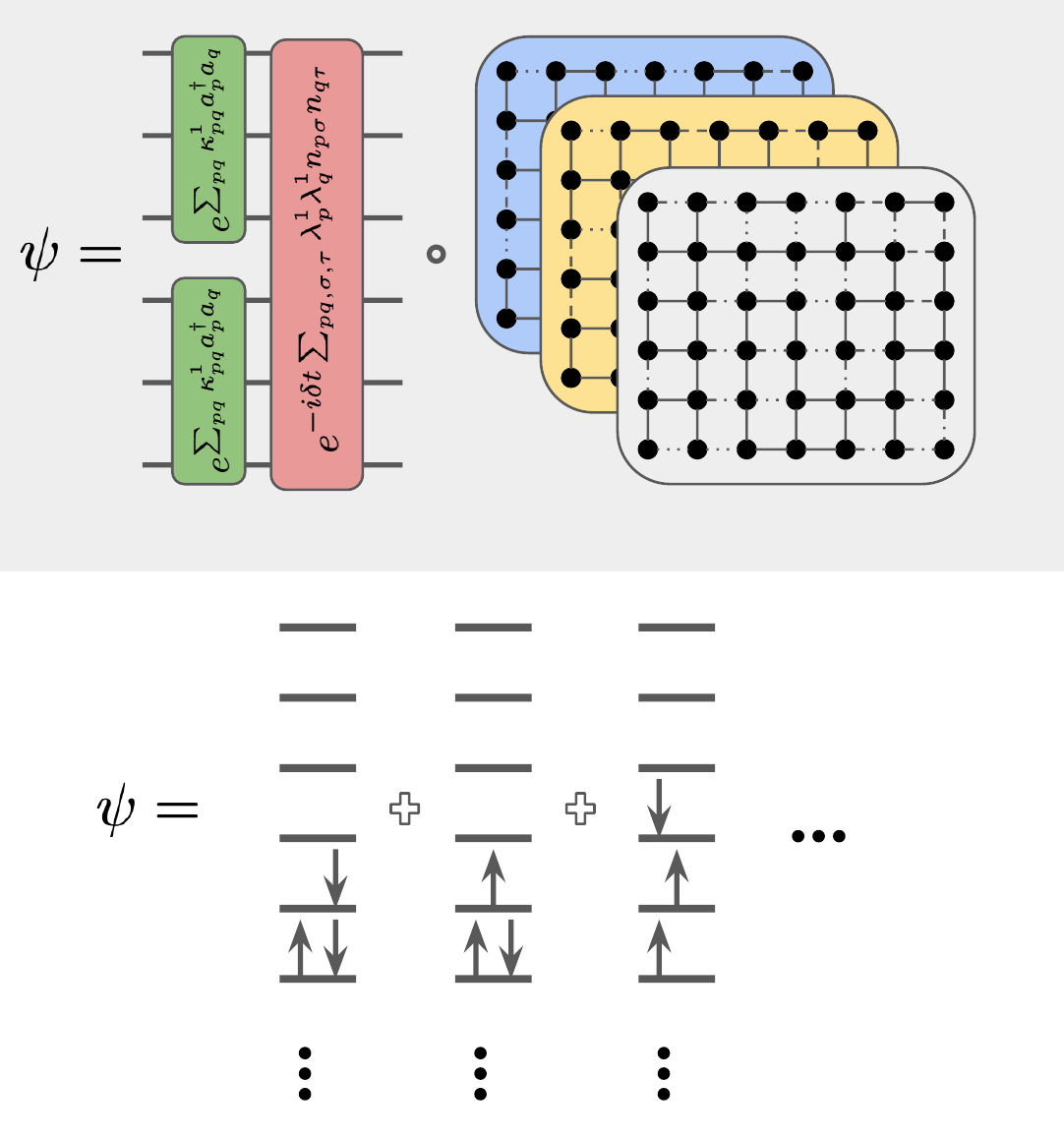}
    \caption{ Contrasting the state vs stabilizer and rotation viewpoint for quasi-degenerate systems.  The top shows a representation of an electronic state built from rotations applied to a stabilizer state, while the bottom depicts a traditional configuration based view.  Electronic catalysts often exhibit quasi-degeneracy, making them challenging to treat with single reference methods.  The stabilizer formalism from quantum error correction, combined with efficient single particle rotations, may offer a compact way to both simulate and analyze such situations, as the use of degeneracy to protect quantum states is a core principle of quantum error correction. Graph state representations of stabilizer states may offer connections to chemical bonding and electron correlation theory, all while being computationally efficient to simulate and analyze classically.
    \label{fig:QEC_Connnection}
    }
\end{figure}

The development of quantum error correction was also closely linked to the development of simulation for a particular class of quantum states, known as stabilizer states.  These states can exhibit essentially maximal entanglement across a system, yet are incredibly efficient to simulate classically, often scaling like $N^2$ or $N^3$ in the number of orbitals~\cite{gottesman1998heisenberg,aaronson2004improved}.  While they are discrete in nature, requiring some novel optimization techniques like reinforcement learning, if coupled with simple orbital rotations, these could form yet another powerful ansatz for efficient classical electronic structure methods informed by quantum information that has yet to be exploited.  We depict this connection in Fig.~\ref{fig:QEC_Connnection}.

While the existence of a powerful ansatz that has not yet been explored is perhaps strong enough motivation by itself, the link to quantum error correction provides even more fuel for this interesting avenue of inquiry.  Many of the most challenging bio-metallic enzymes, such as nitrogenase, are characterized by an active site that is difficult to describe with traditional methods due to static correlation related to quasi-degeneracy of electronic configurations.  Quantum error correction takes advantage of strong entanglement and engineered degeneracy as a mechanism for protecting information.  If nature were to follow a similar path, perhaps the electronic near-degeneracy and resulting entanglement in these systems could be an avenue for protecting a coherent reaction pathway similar to the approach of quantum error correction.  In addition, many modern quantum error correcting codes take advantage of topological protection, and topological effects have been studied in the context of electronic degeneracies near conical intersection~\cite{yarkony1996diabolical}.  While natural systems do not have access to an optimal decoder to remove excess entropy, the notion of self-correcting memories where systems exhibit such behavior more naturally draws an enticing connection~\cite{brown2016quantum}.  In such a case, the stabilizer formalism for representing these ground states would be especially apt, efficiently manipulating untold numbers of determinants through non-obvious symmetries~\cite{garcia2017geometry}.  This could offer an electronic (non-point group) symmetry perspective view on important intermediate or catalytic electronic states.  Such statements are presently at the level of conjecture, but it represents another example of a potential way in which quantum computation and chemistry may unite.   We depict this perspective in Fig.~\ref{fig:QEC_Connnection}.

\section{Outlook}

As quantum technology advances, so does our understanding of what any computational device can accomplish.  In this perspective, we have explored the ways in which results from quantum computer science may impact our view and approach towards computational chemistry.  On one hand, quantum computers are believed to offer an exponential advantage over classical computers in direct simulation of some quantum chemical theories, making certain tasks that previously seemed impossible into ones that should be relatively routine.  On the other hand, we saw that even quantum computers have limits, crystallized through the no-fast-forwarding theorem and strong results on undecidability of physical processes with quantum computers.

In highlighting the ways some of these speed limits are broken, we were led to a framework where learning from natural data is fundamentally different from rote computation.  This viewpoint captures the ability of reduced theories to offer meaningful predictions further than the timescales of experiments, and also offers a strategy for dealing with some of the hardest problems in chemical theory.  

It has now been decisively shown that classical models empowered by data from quantum computation (including both nature and engineered computers) are more powerful than traditional computation without data, assuming only that quantum computers may perform some tasks faster than classical computers.  While data from quantum devices will play an important role in their future interactions with chemistry, the ability to feed quantum data directly into quantum computers offers more power still.  The interplay of quantum computers with more advanced quantum sensors may offer untold possibilities.  Finally, we expect the interplay between quantum information theory and chemical theory to continue, where import of ideas from domains like digital quantum error correction may open new avenues of research.

In looking forward, we see a bright future for the ways in which quantum technology may advance the study of chemistry.  While it is true that the direct simulation abilities of quantum computers will offer amazing value, we have argued here that this is just the tip of the iceberg.  Our ability to address some of the hardest problems in chemistry and nature through data and learning will only be bolstered by stronger quantum technology, and that future is on its way.

\section{Acknowledgements}
We thank Nathan Wiebe for helpful discussions and feedback on the draft.

\bibliographystyle{apsrev4-1_with_title}
\bibliography{references}

\end{document}